\documentclass[
    reprint,
    amsmath,amssymb,
    aps,
    pra
]{revtex4-2}

\usepackage[utf8]{inputenc}

\usepackage{hyperref}
\usepackage{graphicx}    % figuras
\usepackage{placeins}   % já ok
\usepackage{dcolumn}     % colunas decimais em tabelas
\usepackage{bm}          % negrito em matemática
\usepackage{xcolor}      % cores
\usepackage{braket}      % notação de Dirac
\usepackage{physics}     % derivadas, operadores, etc.
\usepackage{tikz}        % diagramas
\usetikzlibrary{matrix, positioning}
\usepackage{soul}        % destaque de texto
\usepackage{booktabs}    % linhas de tabela melhoradas
\usepackage{stfloats}    % melhora o controle de figure*
\usepackage{placeins}    % \FloatBarrier
\usepackage{lipsum}      % texto de exemplo (opcional)
\usepackage{caption}
\usepackage{newtxtext, newtxmath}
\begin{document}

\preprint{APS/123-QED}

% \title{Shaping Dynamics Through Memory: A Study of Reservoir Profiles in Open Quantum Systems}
\title{Non-markovian reservoir profile effects on dynamics of light within a single waveguide}

\author{J. R. Silva}
\email{jefferson.rocha@fis.ufal.br}
\author{C. Antunis B. S. Santos}
\email{carlos.bonfim@fis.ufal.br}

\affiliation{Instituto de F\'isica, Universidade Federal de Alagoas, Macei\'o, Brasil}

\date{\today}

\begin{abstract}
    In this work, we investigate how different reservoir memory profiles influence the dynamical evolution of a single waveguide coupled to an external environment. We compare three representative memory kernels: Lorentzian, Gaussian and Uniform, highlighting their distinct spatial correlations and their impact on system behavior. We compute the transmission amplitude, transparency properties, as well as long-time behavior of the system under each memory model. To quantify deviations from Markovian dynamics, we employ a non-Markovianity measure based on information backflow, allowing a direct comparison between the structured reservoirs and the Markovian limit. Our results reveal clear signatures of memoryless-induced modifications in the transmission spectrum and demonstrate how specific reservoir profiles enhance or suppress non-Markovian effects.
\end{abstract}

\maketitle

%\tableofcontents

\section{Introduction}

Open quantum systems provide a natural framework for describing the dynamics of physical platforms that interact with external environments. In photonic and condensed-matter settings, such environments are often modeled as reservoirs coupled to a localized mode or guided field \cite{Ferreira2021, Longhi2006}. These interactions allow for the exchange of energy and information, leading to complex non-Markovian dynamics and quantum correlations \cite{Budini2022, Berrada2025}. The properties of the reservoir as its spectral density, spatial correlations, and characteristic decay, play a decisive role in shaping system evolution \cite{Galve2016, Gaidi2026}. Depending on the structure of these correlations, the dynamics may range from fully Markovian, where memory effects are negligible, to strongly non-Markovian regimes in which information can effectively flow back from the environment to the system. Understanding how different reservoir profiles influence observable quantities is therefore essential for both fundamental studies and technological applications such as light transport, dissipative engineering, and quantum information processing \cite{Bylicka2014, Li2022, Shirai2021, ElAllati2024}.

In waveguide–reservoir architectures, environmental memory manifests itself explicitly through convolution equations that couple the field amplitude at a given position to its entire spatial interaction history \cite{Vega2017, Tufarelli2014}. This nonlocal dependence highlights how reservoir structure governs spatial propagation rather than simply modifying attenuation strengths. Different reservoir kernels capture distinct physical mechanisms: Markovian reservoirs correspond to memoryless exponential correlations and serve as a reference model for irreversible propagation \cite{Gardiner1985}; Lorentzian reservoirs encode finite spatial correlation lengths associated with cavity-filtered or resonant spectral densities \cite{Garraway1997}; Gaussian and Uniform kernels describe structured or broadband environments, often relevant in engineered photonic media or environments shaped by fabrication constraints \cite{Vasile2011, Calajo2019}. Although each of these kernels has been studied individually in various contexts, a systematic comparison of their impact on propagation within a single, controlled scenario remains largely unexplored \cite{Breuer2016, Liu2011}.

Consequently, elucidating the specific dynamical signatures imprinted by these distinct reservoir classes is a key step toward effective reservoir engineering. The ability to distinguish between effects arising from a sharp spectral cutoff versus a broad interaction bandwidth allows for more precise control over light transport. By mapping the relationship between the functional form of the memory kernel and the emergence of phenomena such as transparency or bound states, providing a roadmap for design photonic structures that can selectively suppress or enhance dissipation based on the available environmental spectral density.

This work presents an unified and quantitative analysis of how three representative reservoir memories affect the spatial evolution of a single-mode waveguide coupled to an external environment. Initially, we derive the analytical solution starting from the system's Hamiltonian. Subsequently, we analyze the spatial transmission profiles and the conditions for the onset of transparency. Finally, we apply a measure of non-Markovianity to quantify information backflow and establish a memory hierarchy among the reservoirs.

\section{Hamiltonian and Heisenberg equations solution}

The system describes an optical waveguide $G$ coupled to a structured reservoir composed of modes, as illustrated in figure \ref{model}.

\begin{center}
  \includegraphics[width=0.48\textwidth]{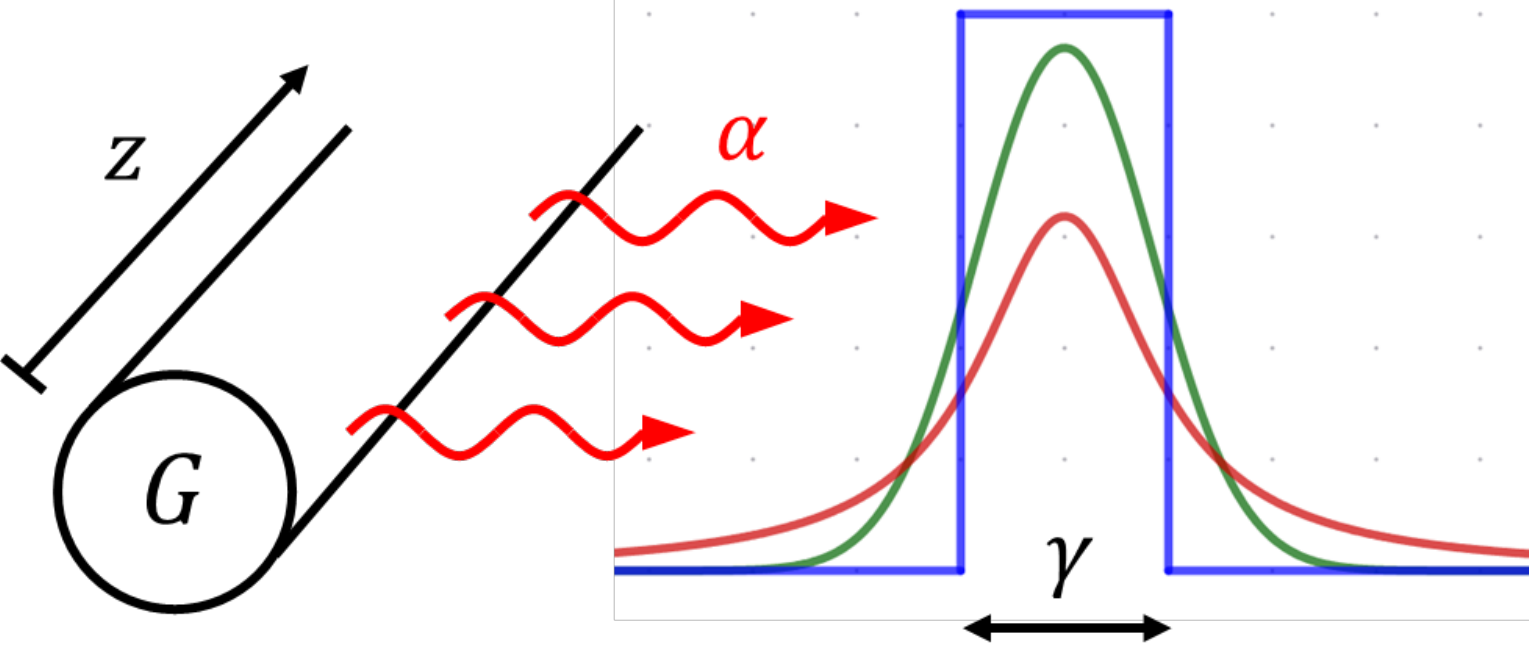}
  % legenda manual (não é ideal para referências automáticas)
  \captionof{figure}{G-waveguide coupled to an environment by a parameter $\alpha$ that can exhibit different mode distributions with a fixed width $\gamma$.}
  \label{model}
\end{center}

The Hamiltonian $H$ consists of: The operator $a$ cancels an excitation in the guided mode, while $r_n$ cancels an excitation in the $n$th reservoir mode. The parameter $\beta$ represents the propagation constant of the isolated guide, while $\beta_n$ denotes the propagation constant of each reservoir mode.

\begin{equation}
    H=\beta a^{\dagger}a+\sum_n \left[\beta_n r_n^{\dagger}r_n+\alpha(\beta_n) (r_n^{\dagger}a+r_na^{\dagger})\right]
\end{equation}

Here, $\alpha(\beta_n)$ represents the spectral coupling function, encoding how strongly the guided mode interacts with each mode of the environment. The interaction term $r_n^{\dagger} a + r_n a^{\dagger}$ facilitates the coherent exchange of energy while discarding rapidly oscillating counter-rotating terms \cite{bloch1940magnetic, jaynes2005comparison}. Consequently, the nature of the reservoir, specifically the spectral density defined by the distribution of $\alpha(\beta_n)$, determines the dynamical regime, which can induce significant non-Markovian features.

Overall, this Hamiltonian captures the essential physics of a single optical mode interacting with a continuum of modes whose spectral density is determined by the dispersion relation of $\beta_n$ and the form of $\alpha(\beta_n)$.

In the Heisenberg representation, the evolution of the operators $a$ and $r_n$ is given by
\begin{equation}\label{a}
    \frac{da}{dz} = -i[a, H] = -i\beta\, a - i \sum_n \alpha(\beta_n)\, r_n  
\end{equation}
and
\begin{align}\label{rn}
    \frac{d}{dz}r_n = -i[r_n, H] = -i\beta_n\, r_n - i \alpha(\beta_n)\, a.
\end{align}

By integrating equation \eqref{rn} and substituting $r_n$ into equation \eqref{a}, we obtain

\begin{equation}\label{da}
    \begin{split}
        \frac{d}{dz}{a} =& -i\beta a
        - \int_0^z \left[\sum_n \left[\alpha(\beta_n)\right]^2e^{-i\beta_n (z-z')}\right]a(z')dz'\\
       &- i \sum_n \alpha(\beta_n) e^{-i\beta_n z} r_n(0).        
    \end{split}
\end{equation}

In the limit where $n$ is a continuous index, we take $\sum_n\left[\alpha(\beta_n)\right]^2\xrightarrow{}\alpha^2\int d\beta' \rho(\beta')$, where $\rho$ is a density function that distributes the reservoir modes. Thus, we construct an autocorrelation function $m(z)$ of the form
\begin{equation}
    m(z)=\alpha^2\int \rho(\beta')e^{-i\beta' z}d\beta',
\end{equation}

as indicated by the Wiener-Khichin theorem \cite{goodman2015statistical}. To evaluate the different possible regimes generated by each distribution, consider Table \ref{t1} with the particular choice of three probability distributions with their autocorrelations and Full Widths at Half Maximum (FWHM), the latter being the key parameter to match the communication window between the photons propagating in the waveguides and the reservoir modes.

\begin{widetext}
\begin{center}
\begin{tabular}{c|c|c|c}
\textbf{Distribution} &
\textbf{PDF} &
\textbf{Autocorrelation} &
\textbf{FWHM}\\
\midrule \midrule

Lorentzian &
$\displaystyle
\frac{1}{\pi}\frac{\Gamma}{(x-\beta_c)^2+\Gamma^2}
$ &
$\displaystyle 
\alpha^2e^{-i\beta_c z}\,e^{-\Gamma z}
$ & $2\Gamma$
\\
\midrule

Gaussian &
$\displaystyle 
\frac{1}{\sigma\sqrt{2\pi}}
\exp\left[-\frac{(x-\beta_c)^2}{2\sigma^2}\right]
$ &
$\alpha^2\displaystyle 
e^{-i\beta_c z}\,e^{-\sigma^2z^2/2}
$ & $\sqrt{\ln{256}}\sigma$
\\
\midrule

Uniform &
$\displaystyle
\frac{1}{\gamma}\, \Theta\left(\frac{\gamma}{2} - |x - \beta_c|\right)
$ &
$\alpha^2\displaystyle 
e^{-i\beta_c z}\text{sinc}(\gamma z/2)
$ & $\gamma$
\\
%\midrule

%Ohmic &
%$\displaystyle
%\frac{x}{\beta^2}e^{-x/\beta}\Theta(x)
%$ &
%$\displaystyle 
%\frac{1}{(1+i\beta z)^2}
%$ & $\gamma\approx2.44638603703\beta$
%\\
\end{tabular}
\captionof{table}{Known probability density functions centered on $\beta_c$ with their autocorrelations and width at half-height.}
\label{t1}
\end{center}
\end{widetext}

We rewrite equation \eqref{da} in the form
\begin{equation}\label{qn}
    \frac{d}{dz}a=-i\beta a-m*a+q.n.
\end{equation}
$q.n.$ in equation \eqref{qn} refers to quantum noise terms \cite{clerk2010introduction}, which can be neglected when choosing states where the reservoir starts the dynamics empty.

The solution of equation \eqref{qn} for $a(z)$ has the known form $a(z)=e^{-i\beta z}f(z)a(0)$ \cite{de2017dynamics}, which, under states where the reservoir is initially empty ($\ket{G}\ket{0}_R$) we can rewrite \eqref{qn}
\begin{equation}\label{dfmf}
    \frac{d}{dz}f(z)=-(\tilde{m}*f)(z),
\end{equation}
where $\tilde{m}(z)=m(z)e^{i\beta z}$. Considering non-markovian cases, such as the mentioned in the Table \ref{t1}, the function $f$ is subject to the boundary condition $f(0)=1$ and, as a consequence of the integro-differential nature of the equation, we obtain $\frac{d}{dz}f|_{z=0}=0$.

% Physically, the term $|f\,|^2$ is the probability of incident photons being found in the guide and its rate,
% \begin{equation}
% \begin{split}
%     \frac{d}{dz}|f(z)|^2&=2\text{Re}\left\{f^*(z)\frac{d}{dz}f(z)\right\}\\
%     &=-2\text{Re}\left\{f^*(z)(\tilde{m}*f)(z)\right\},
% \end{split}
% \end{equation}
% is given by the flux rate or photon flux to the reservoir. When the term $2\text{Re}{[f^*(\tilde{m}*f)]}$ is positive, the waveguide acts as a sink, resulting in probability leakage. Conversely, if it is negative, the waveguide acts as a source, with the reservoir injecting probability back into the system (Revival).

\section{Transmission}

In the context of wave physics, optics, and open quantum systems, transmission is a dimensionless quantity that quantifies the fraction of power, intensity, or probability that manages to traverse a system or medium and reach the other side. Considering a state $\ket{G}$ for the size guide $z$, transmission can be expressed as

\begin{equation}
    T(z)=\frac{\bra{G,0_R}a^\dagger(z)a(z)\ket{G,0_R}}{\bra{G,0_R}a^\dagger(0)a(0)\ket{G,0_R}}=|f(z)|^2
\end{equation}

showing the physical interpretation of $|f|^2$, which is the probability of incident photons being found in the guide. By \eqref{dfmf}, one could derive the dynamic of this quantity, as shown in the \eqref{flxrtT}
\begin{equation}\label{flxrtT}
\begin{split}
    \frac{d}{dz}|f(z)|^2&=2\text{Re}\left\{f^*(z)\frac{d}{dz}f(z)\right\}\\
    &=-2\text{Re}\left\{f^*(z)(\tilde{m}*f)(z)\right\},
\end{split}
\end{equation}
expressing the flux rate or photon flux to the reservoir.  When the term $2\text{Re}{[f^*(\tilde{m}*f)]}$ is positive, the waveguide acts as a sink, resulting in probability leakage. Conversely, if it is negative, the waveguide acts as a source, arising a revival on the guide where the reservoir injects probability back into the system.

We initially evaluated four distinct configurations. The first three correspond to open systems coupled to reservoirs described in Table \ref{t1} under the resonance condition $\beta_c=\beta$. Importantly, this choice ensures that the spectral width is much smaller than the central frequency ($\gamma \ll \beta_c$), which justifies the approximation of extending the integration limits to analytically derive the autocorrelation functions. Finally, we considered a limiting case common to all reservoirs where $\text{FWHM}\equiv\gamma=0$. This condition is equivalent to the reservoir acting as a coupled single-mode waveguide, yielding $T(z)=\cos^2(\alpha z)$, as shown in Figure \ref{Tr}.

\begin{center}
  \includegraphics[width=0.48\textwidth]{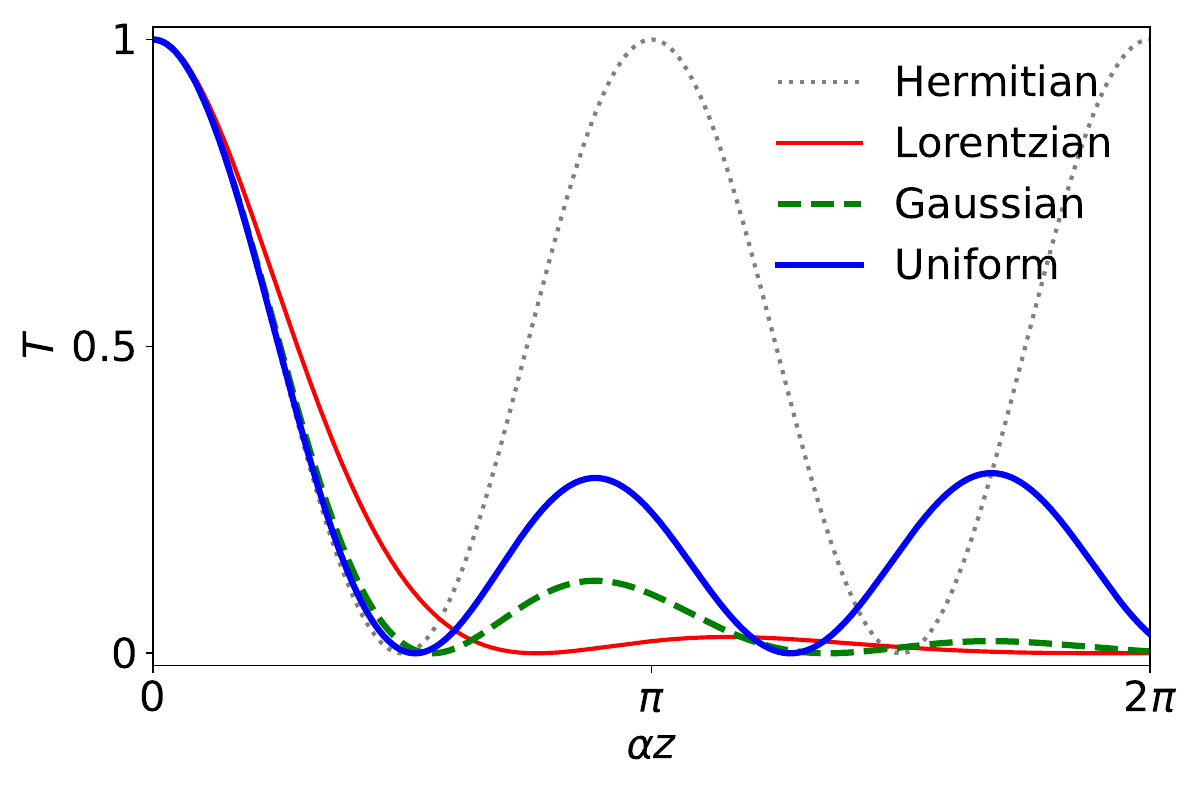}
  % legenda manual (não é ideal para referências automáticas)
  \captionof{figure}{Transmission with $\gamma=2\alpha$ for three reservoirs highlighted with solid and dashed lines and the Hermitian situation where $\gamma=0$ with a dotted line.}
  \label{Tr}
\end{center}

Among the uniform, Gaussian, and Lorentzian distributions, the Uniform distribution is expected to exhibit the most pronounced transmission revivals. To analyze the consistency of this result, observe that the Fourier transform of a Gaussian is another Gaussian; therefore, the memory decay is very rapid and smooth. The Fourier transform of a Lorentzian, otherwise, is a decaying exponential, and although oscillations may arise under strong detuning, the natural tendency of the memory envelope is to decay monotonically. Finally, the Fourier transform of a Rectangular function is a cardinal sine (sinc function), which possesses strong intrinsic oscillations and decays slowly as $1/z$. These oscillations in the memory function signify that the reservoir returns information to the system rhythmically and intensely before complete dissipation.

The physics of these revivals is intimately linked to the spectral smoothness. In the Lorentzian and Gaussian cases, the distributions possess smooth tails extending to infinity. A photon entering the reservoir always finds available modes, facilitating the forgetting process or irreversible dissipation. In contrast, the uniform case features abrupt edges (hard cutoffs) where the spectrum ends suddenly. When a photon attempts to decay into propagation constants outside the uniform reservoir's band, it finds no available modes. These edges effectively act as walls in wavenumber space; the photon strikes the band edge and reflects back into the waveguide. 

Despite knowing the autocorrelation explicitly, it is not possible to write a closed-form solution for the function $f$ in terms of known functions for an arbitrary reservoir. In the power series expansion, using the boundary condition $f_0=1$ and additionally $f_1=0$ as previously mentioned, the remaining coefficients follow from the expansion of equation \eqref{dfmf}:
\begin{equation}    f_{n+1}=-\frac{1}{(n+1)^2}\sum_{j=0}^n\frac{m_j}{\binom{n}{j}} f_{n-j},\quad n>1.\end{equation}We can write $f(z)$ in terms of the autocorrelation function expansion coefficients such that:

\begin{equation}    f(z)=1-\frac{m_0}{2}z^2-\frac{m_1}{6}z^3+\frac{m_0^2-2m_2}{24}z^4+\dots.\end{equation}

For the three previously selected reservoirs, the solution expansion is tabulated below.
\begin{center}
\small
\begin{tabular}{c|c}
$\tilde{\mathbf{m}}(z)$    & \textbf{Series expansion}   \\
\midrule\midrule
\multicolumn{2}{c}{Lorentzian} \\
\midrule
$\alpha^2e^{-\gamma z/2}$ & $1-\frac{\alpha^2}{2}z^2+\alpha^2\frac{\gamma}{12}z^3+\frac{\alpha^2}{24}\left(\alpha^2-\frac{\gamma^2}{4}\right)z^4+\dots$ \\
\midrule
\multicolumn{2}{c}{Gaussian} \\
\midrule
$\alpha^2\exp[-\frac{\gamma^2}{\ln 65536}z^2]$ & $1-\frac{\alpha^2}{2}z^2+\frac{\alpha^2}{24}\left(\alpha^2+\frac{\gamma^2}{\ln256}\right)z^4+\dots$ \\
\midrule
\multicolumn{2}{c}{Uniform} \\
\midrule
 $\alpha^2\text{sinc}(\gamma z/2)$ & $1-\frac{\alpha^2}{2}z^2+\frac{\alpha^2}{24}\left(\alpha^2+\frac{\gamma^2}{12}\right)z^4+\dots$\\
\end{tabular}
% \begin{tabular}{c|c}
% \midrule

% $\tilde{m}(z)$ & \textbf{Lorentzian} \\
% \midrule 
% $\alpha^2e^{-\gamma z/2}$ & $1-\frac{\alpha^2}{2}z^2+\alpha^2\frac{\gamma}{12}z^3+\frac{\alpha^2}{24}\left(\alpha^2-\frac{\gamma^2}{4}\right)z^4+\dots$ \\
% \midrule 
% $\tilde{m}(z)$ & \textbf{Gaussian} \\
% \midrule
% $\alpha^2\exp[-\frac{\gamma^2}{\ln 65536}z^2]$ & $1-\frac{\alpha^2}{2}z^2+\frac{\alpha^2}{24}\left(\alpha^2+\frac{\gamma^2}{\ln256}\right)z^4+\dots$ \\
% \midrule
% $\tilde{m}(z)$ & \textbf{Uniform} \\
% \midrule
%  $\alpha^2\text{sinc}(\gamma z/2)$ & $1-\frac{\alpha^2}{2}z^2+\frac{\alpha^2}{24}\left(\alpha^2+\frac{\gamma^2}{12}\right)z^4+\dots$\\
% \midrule
% \end{tabular}
\captionof{table}{Expansion of the solution of the equation for each autocorrelation function.}
\label{t2}
\end{center}

Note that the solutions for all proposed reservoirs share expansion terms up to the second order, which explains the similar quadratic decay behavior of the transmissions observed in Figure \ref{Tr}. Furthermore, the cubic term, presented only in the Lorentzian case, indicates higher transmission for shorter propagation distances.

When the FWHM is sufficiently broadened, a Markovian behavior is induced, with $T$ becoming practically exponential, following a Beer-Lambert-type law. A comparative framework for each reservoir is presented in Figure \ref{fM}.

\begin{center}
  \includegraphics[width=0.48\textwidth]{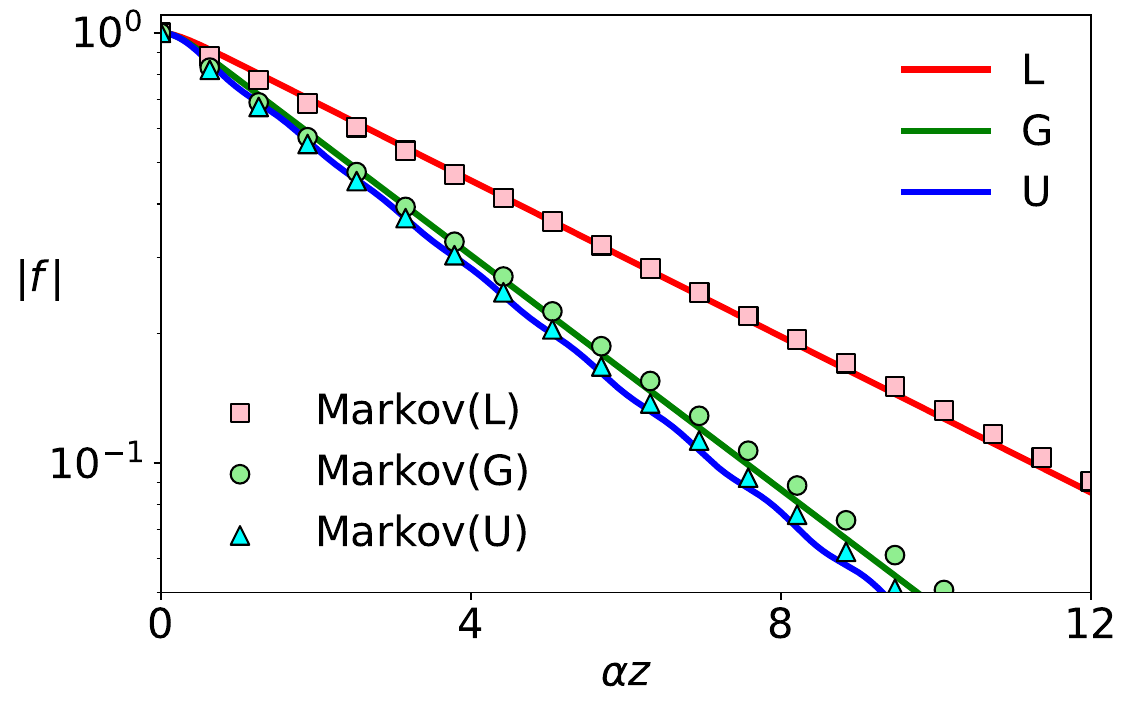}
  \captionof{figure}{function $|f|$ on a logarithmic scale with $\gamma = 10\alpha$ for each reservoir, with the addition of points indicating Markovian decays for comparative purposes.}
  \label{fM}
\end{center}

Decay rates can be obtained by the semidefinite integral of the kernel $\tilde{m}$. For the Lorentz reservoir $\gamma e^{-\gamma z/2}\xrightarrow{}2\delta(z)$, to the Gauss reservoir  $\gamma\,\exp\left[-\frac{\gamma^2}{\ln 65536}z^2\right]\xrightarrow{}\sqrt{\pi \ln 16}\,\,\delta(z)$, and for the uniform reservoir  $\gamma\,\text{sinc}(\gamma z/2)\xrightarrow{}\pi\delta(z).$

The results obtained for $|f|^2$ are consistent with Fermi's Golden Rule \cite{fox2006quantum}, which serves as the macroscopic limit of the problem. In this Markovian regime, the decay rate is directly proportional to the spectral density at resonance, given by $\Gamma \propto 2\pi\alpha^2\rho(\beta)$. By comparing the distributions for a fixed Full Width at Half Maximum (FWHM), we observe that the Uniform distribution yields the most pronounced decay. This occurs because the rectangular profile concentrates the spectral weight entirely within the bandwidth, maximizing $\rho(\beta_c)$. It is followed by the Gaussian case, while the Lorentzian distribution exhibits the slowest decay. The latter behaves this way because its heavy tails distribute a significant portion of the spectral density away from the resonance, resulting in the lowest peak value $\rho(\beta_c)$ among the three configurations.

\section{Transparence}

Optical transparency \cite{guo2009observation} refers to the photon transport capability within a waveguide without significant absorption by the reservoir. In this context, Loss-Induced Transparency (LIT) \cite{beder2024quantum,silva2025anomalous} stands out as a counter-intuitive non-Hermitian effect, where enhancing dissipation in specific channels favors overall transmission. While LIT is traditionally associated with spatial losses, we present in Figure \ref{Trg} a dynamical analogue of this effect: transparency assisted by memory suppression.

\begin{center}
  \includegraphics[width=0.45\textwidth]{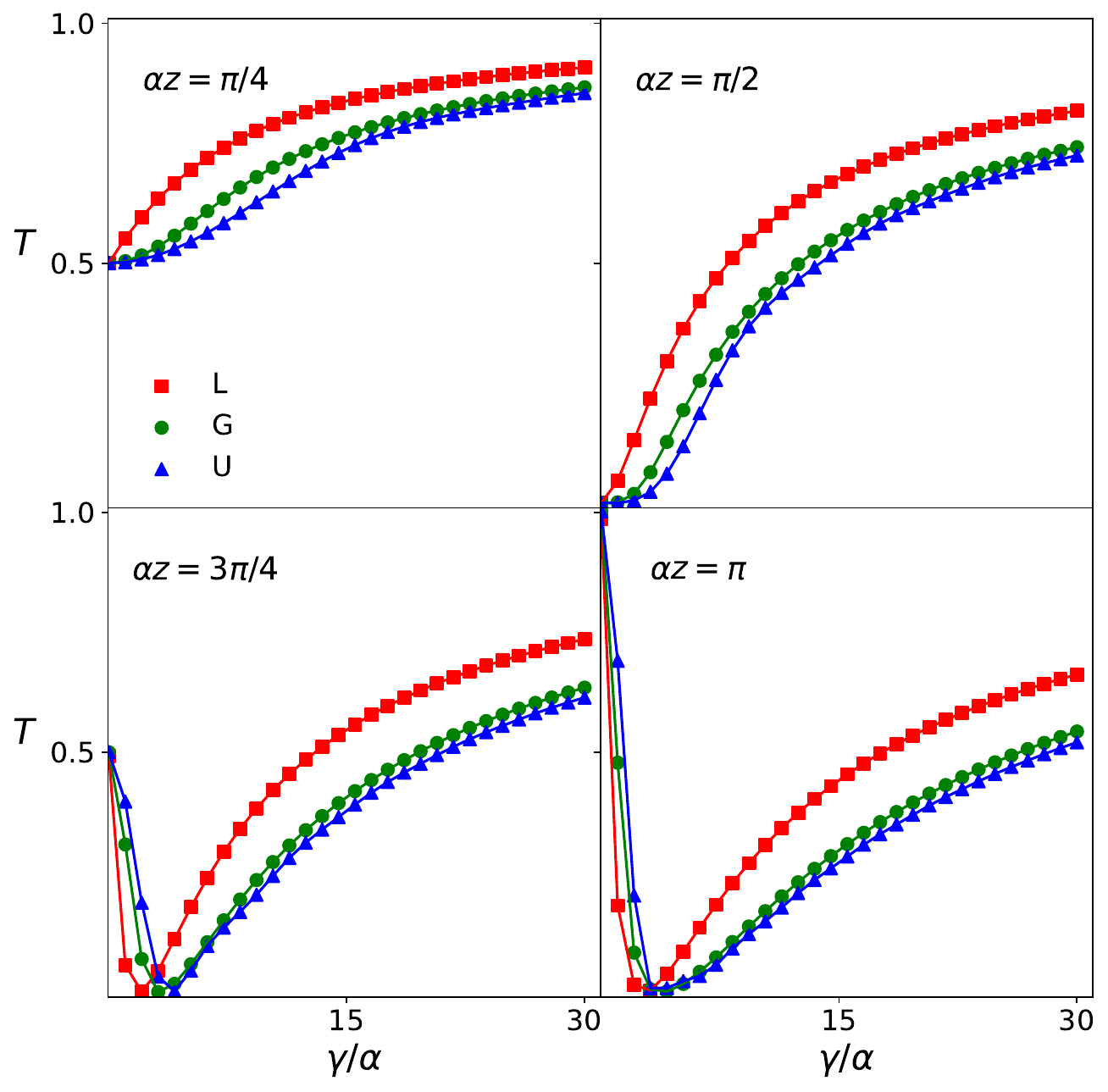}
  % legenda manual (não é ideal para referências automáticas)
  \captionof{figure}{Transmission to the three registers as a function of $\gamma/\alpha$ with $\alpha z=\pi/4,\,\pi/2,\,3\pi/4,\,\pi$.}
  \label{Trg}
\end{center}

We observe that the transmission increases monotonically with the broadening of the reservoir spectral bandwidth in $\alpha z=\pi/4$ e $\alpha z=\pi/2$. Physically, the reservoir fluctuations become too fast compared to the system, thereby dissipating energy inefficiently, leading to enhanced population survival.

The non-monotonic transmission behavior observed at $\alpha z=3\pi/4$ and $\alpha z=\pi$ as a function of the reservoir spectral width is governed by the interplay between two distinct physical mechanisms. In the narrow-bandwidth regime, high transmissivity is sustained by long-range memory effects, where strong coupling coherence facilitates energy backflow (revivals) prior to irreversible dissipation. An initial increase in $\gamma$ degrades this temporal coherence, suppressing the return oscillations and driving the system toward a critical point of maximum opacity. However, as the bandwidth expands beyond this threshold, a spectral dilution regime emerges. In this scenario, the local density of states at the resonance frequency is drastically reduced, effectively weakening the waveguide-reservoir interaction. This inhibits the decay rate and restores system transparency through a decoupling mechanism induced by the dispersal of bath modes.

Comparatively, in the memory regime (small $\gamma$), the Uniform reservoir exhibits the highest transmission due to stronger revival dynamics, followed by the Gaussian, while the Lorentzian profile shows the lowest transmission, proving to be the most dissipative. Conversely, for large $\gamma$, this hierarchy is reversed, and the Lorentzian profile becomes the most transparent. In this limit, the dominant mechanism is the suppression of the spectral density at the resonance peak: a lower available density of states for coupling minimizes photon-reservoir interaction, thereby enhancing waveguide transparency.

\section{Non-Markovianity measure}

To quantify the memory effects of the dynamics, we employ the non-Markovianity measure proposed by Breuer, Laine, and Piilo (BLP measure) \cite{chiriaco2025computable, guarnieri2014quantum, breuer2009measure}, which is based on the variation of distinguishability between quantum states. Although the general definition of this measure requires an optimization over all possible pairs of initial states to detect information backflow, the system studied here, a waveguide in the single-excitation regime interacting with the environment, belongs to the class of amplitude damping channels. For this specific class, it has been analytically demonstrated that the pair of states maximizing the information flow is the orthogonal pair formed by the vacuum state $|0\rangle$ and the excited state $|1\rangle$ \cite{breuer2009measure, breuer2016colloquium}. Consequently, the trace distance becomes exactly equal to the absolute value of the photon amplitude function, $|f(z)|$. Thus, the non-Markovianity measure $\mathcal{N}$ simplifies to the integral of the positive derivative of $|f(z)|$,
\begin{equation}
\mathcal{N}=\int_{dT>0}\frac{d}{dz}|f(z)|\,dz,
\end{equation}
cumulatively accounting for all propagation regions where the rate of change of the amplitude is positive. Physically, this corresponds to the intervals where the decay flow is reversed, leading to a recovery of coherence by the system.

\begin{center}
  \includegraphics[width=0.48\textwidth]{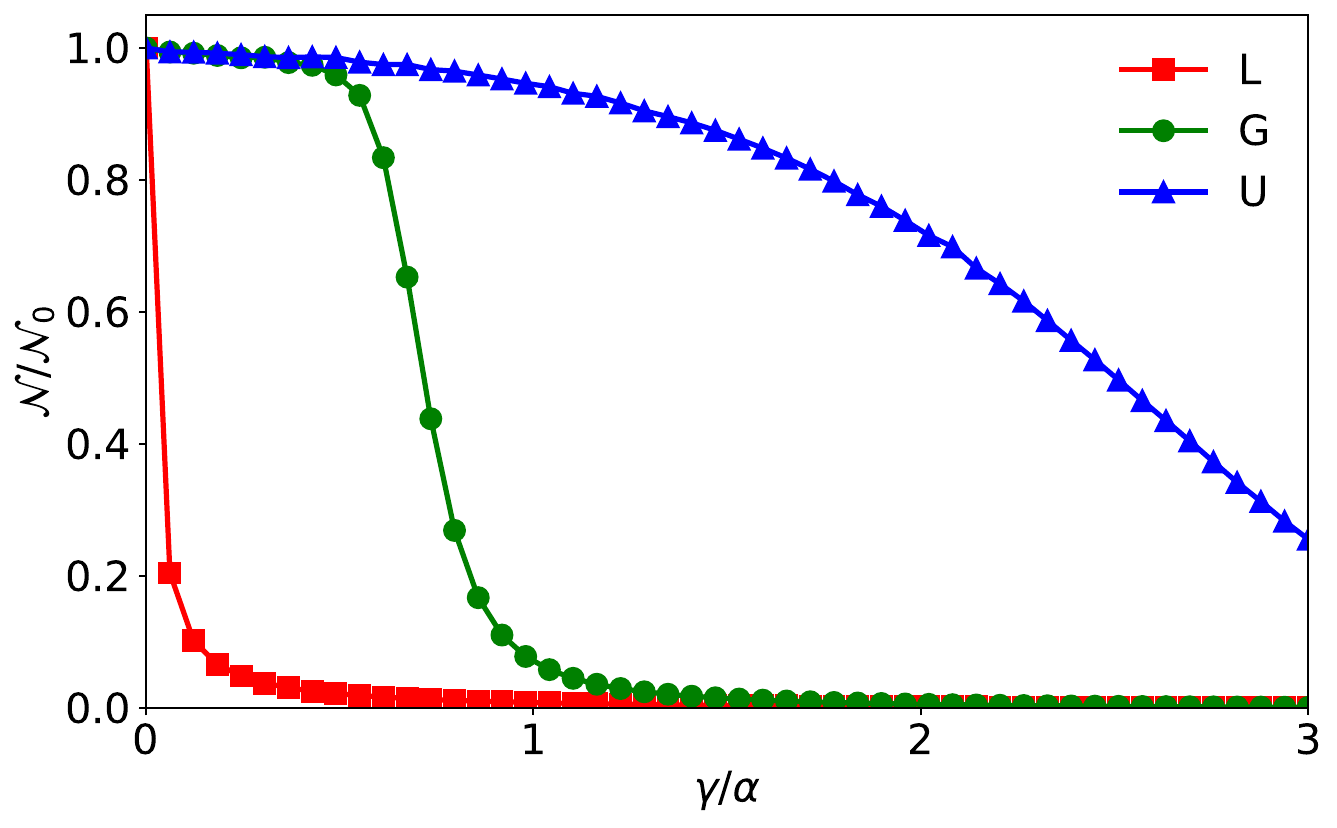}
  % legenda manual (não é ideal para referências automáticas)
  \captionof{figure}{BLP measurement curves for comparative evaluation of memory between three reservoirs.}
  \label{blp}
\end{center}

In the Hermitian limit ($\gamma=0$), where the reservoir effectively behaves as a secondary waveguide, the BLP measure yields $\mathcal{N}_0=\lfloor\alpha L/\pi\rfloor+[1-\Theta(\tan(\alpha L))]|\cos(\alpha L)|$ for a waveguide of length $L$. In the simulation shown in Figure \ref{blp}, we used $\alpha L=100\pi$. All curves in Fig. \ref{blp} exhibit a monotonic decay. Regarding memory effects, we observe that spectral distributions which are more concentrated and possess lighter tails correspond to stronger non-Markovian features in the system-reservoir interaction.

The physical interpretation of the results obtained via the BLP measure can be further elucidated through the pseudo-mode approach \cite{Garraway1997, Mazzola2009, Pleasance2020}, adapted here to the spatial domain. In this framework, the reservoir's density of propagation constants is analytically mapped onto a set of fictitious auxiliary modes interacting with the main waveguide. A Lorentzian distribution of propagation constants, characterized by long, smooth tails in reciprocal space, mathematically corresponds to a single pole in the complex plane. This is equivalent to coupling the waveguide to a single lossy pseudo-mode, a simple structure that favors purely exponential spatial decay, resulting in essentially Markovian dynamics along the $z$-axis.Conversely, distributions with compact support, such as the uniform distribution, cannot be represented by a single simple pole. The presence of sharp cutoffs in the allowed propagation constants introduces singularities that effectively require a superposition of multiple pseudo-modes to describe the spatial interaction. The BLP measure thus captures the interference effects between these modes during propagation, creating a coherent framework that allows to evaluate the revivals of photon amplitude on the main waveguide at subsequent positions, characterizing spatial information backflow.

%\color{blue}
%O caso lorentziano é especial, pois pode ser interpretado como um problema de dinâmica de dois guias acoplados, onde o segundo guia atua como memória do guia de interesse. A propagação dos fótons nesse segundo guia, ou pseudo-guia, segue como se houve-se fuja markoviana com taxa $\gamma$ e, eventualmente, a luz pode ser readmitida pelo guia $G$.
%\color{black}

\section{Conclusions}

In this work, we investigated photon propagation dynamics in waveguides coupled to structured reservoirs, demonstrating that the environment's spectral density profile plays a decisive role in the transition between dissipative and memory regimes. A comparative analysis of Lorentzian, Gaussian, and Uniform distributions, subject to the same Full Width at Half Maximum constraint, revealed fundamentally distinct interaction mechanisms.

We confirmed that the Lorentzian distribution, due to its simple pole analytic structure, effectively acts as a single lossy pseudo-mode. This spectral topology favors irreversible information leakage, resulting in predominantly Markovian dynamics. Although the spectral dilution caused by its infinite tails may lead to slower decay rates in specific regimes, the absence of defined spectral boundaries prevents efficient long-term population preservation. In contrast, distributions with compact support or fast-decaying edges exhibited rich non-Markovian phenomenology. By applying the simplified BLP measure for amplitude damping channels, we quantified the information backflow from the reservoir to the system. The Uniform distribution yielded the highest non-Markovianity values; the sharp discontinuities in its density of states act as reflective barriers in reciprocal space, generating intense transmission oscillations and enabling the formation of quasi-bound states and radiation trapping. Finally, the Gaussian distribution proved to be a robust physical intermediate, offering controlled coherence revivals and transparency windows without the numerical instabilities associated with the singularities of the rectangular profile.

\begin{acknowledgments}
The authors acknowledge the financial support of CNPq (Conselho Nacional de Desenvolvimento Cient\'ifico e Tecnol\'ogico), CAPES (Coordenação de Aperfeiçoamento de Pessoal de Nível Superior) and FAPEAL (Fundação de Amparo à Pesquisa do Estado de Alagoas). 
\end{acknowledgments}

\appendix

%\nocite{*}
\bibliography{apssamp}% Produces the bibliography via BibTeX.

\end{document}